\documentclass[aps,prd,twocolumn,showpacs,preprintnumbers,nofootinbib,amsmath,amssymb]{revtex4}

\usepackage{graphicx}
\usepackage{bm}
\usepackage{amsmath}
\usepackage[normalem]{ulem}
\usepackage{multirow}

\newcommand{\beq}{\begin{eqnarray}}
\newcommand{\eeq}{\end{eqnarray}}

\newcommand{\tr}{\ensuremath{\mathrm{Tr}}}

\def\spose#1{\hbox to 0pt{#1\hss}}
\def\ltapprox{\mathrel{\spose{\lower 3pt\hbox{$\mathchar"218$}}
 \raise 2.0pt\hbox{$\mathchar"13C$}}}

\begin{document}

\title{
Phase diagram of QCD in a magnetic background
}

\author{Massimo D'Elia}
\email{massimo.delia@unipi.it}
\affiliation{
Dipartimento di Fisica dell'Universit\`a
di Pisa and INFN - Sezione di Pisa,\\ Largo Pontecorvo 3, I-56127 Pisa, Italy}

\author{Lorenzo Maio}
\email{lorenzo.maio@phd.unipi.it}
\affiliation{
Dipartimento di Fisica dell'Universit\`a
di Pisa and INFN - Sezione di Pisa,\\ Largo Pontecorvo 3, I-56127 Pisa, Italy}

\author{Francesco Sanfilippo}
\email{francesco.sanfilippo@infn.it}
\affiliation{INFN - Sezione di Roma Tre,\\ Via della Vasca Navale 84, I-00146 Rome, Italy}

\author{Alfredo Stanzione}
\email{alfredo.stanzione@sissa.it}
\affiliation{SISSA, Via Bonomea 265, 34136, Trieste, Italy}

\date{\today}

\begin{abstract}
We provide numerical evidence that the thermal QCD crossover 
turns into a first order transition in the presence
of large enough magnetic background fields. The critical
endpoint is found to be located between $eB = 4$~GeV$^2$
(where the pseudocritical temperature is 
$T_c = (98 \pm 3)$~MeV)
and $eB = 9$~GeV$^2$ (where the critical temperature is
$T_c = (63 \pm 5)$~MeV).
Results are based on the analysis of quark condensates
and number susceptibilities, determined by lattice simulations of
$N_f = 2+1$ QCD at the physical point, discretized with three
different lattice spacings, $a = 0.114, 0.086$ and $0.057$~fm, 
via rooted stout staggered fermions and a Symanzik tree level improved 
pure gauge action. We also present preliminary results regarding
the confining properties of the thermal theory,
suggesting that they could change drastically going 
across the phase transition.
\end{abstract}

\maketitle

\section{Introduction}
\label{intro}

The investigation of QCD properties 
in a magnetic background field has been the subject of 
various studies in the last few years, see, e.g., 
Refs.~\cite{lecnotmag,Andersen:2014xxa,Miransky:2015ava} for recent reviews.
Part of the interest is directly related to phenomenology:
strong background fields are expected in 
non-central heavy ion collisions~\cite{hi1, hi2, hi3, hi4, tuchin,
Holliday:2016lbx}, in astrophysical objects like 
magnetars~\cite{magnetars}, and might have been produced 
during the cosmological electroweak phase
transition~\cite{vacha,grarub}, thus influencing the subsequent 
evolution of the Universe, including the 
cosmological QCD transition.
Lattice QCD simulations
have been essential to advance 
knowledge in this field, given also the fact
that, unlike the case of a baryon chemical
potential, no technical problem hinders the application
of standard Monte-Carlo techniques for the computation of the 
QCD path-integral in a magnetic background.

One of the most relevant aspects regards 
the influence of the magnetic field on the 
QCD phase diagram. 
Early lattice studies of $N_f = 2$ QCD, 
adopting standard staggered fermions 
and {heavier-than-physical} quark masses,
showed a slightly 
increasing behavior of the crossover
temperature $T_c$ as a function of 
the magnetic field $B$~\cite{demusa,Ilgenfritz:2012fw}.
That was however not confirmed by 
an investigation of $N_f = 2+1$ QCD at the physical point
discretized via improved staggered fermion, 
showing instead an appreciable decrease of
$T_c$, of the order of 10-20\%, for
magnetic fields going up to $e B \sim$ 1 GeV$^2$~\cite{reg0},
a behavior confirmed also by later lattice 
studies~\cite{Bornyakov:2013eya}.
The reason for the discrepancy of early results has been
clarified by later studies: it should be ascribed 
to lattice artefacts~\cite{htding0}, while the 
decreasing behavior of 
$T_c$ as a function of $B$ is observed
also for larger than physical pion 
masses~\cite{gabbiano,Endrodi:2019zrl}.

One important aspect, generally confirmed by lattice simulations,
is the strengthening of the QCD crossover as the magnetic 
field is increased, which points to the possibility that
it could turn into a real phase transition for large enough 
$B$. Available predictions, based on the extrapolation of lattice results 
and on the numerical study of effective models, 
suggests that this could happen for $eB$ of the order of 
10~GeV$^2$~\cite{Endrodi:2015oba}. 
That could have significant implications for the physics 
of the Early Universe, regarding in particular 
the consequences of a first order cosmological 
QCD transition~\cite{Witten:1984rs,Applegate:1985qt}.
As a matter of fact, a direct observation of this phenomenon has been 
reported only for discretizations adopting unimproved
staggered quarks~\cite{htding0}, for which however one also observes 
that $T_c$ increases (instead of decreasing) with $B$.

The main purpose of the present study is to push
forward our knowledge on this topic, by exploring 
finite temperature $N_f = 2+1$ QCD with physical 
quark masses at unprecedented values of the magnetic 
field, trying also to keep control on UV cutoff effects.
In order to do that, we will consider a stout improved staggered
discretization of the theory and two different values of 
the magnetic field, $eB \simeq 4$ and 9~GeV$^2$, trying 
to keep control on discretization effects by exploring
three different lattice spacings, $a = 0.057, 0.086$ and $0.114$~fm.
Anticipating part of the final results, we will provide 
evidence that the QCD transition is first order for 
$eB = 9$~GeV$^2$ and that the critical temperature, for
that value of the magnetic field, goes down to values
below 70~MeV. We will also present a preliminary
investigation of the confining properties of the theory,
suggesting that they could change drastically going across
the phase transition.

Even if our investigation is not affected
by any technical obstruction, such as a sign problem,
it is anyway extremely challenging
from a numerical point of view. 
On one hand, the need for large magnetic fields 
requires correspondingly fine lattice spacings, of the order 
or below 0.1~fm. 
On the other hand, given the fact that the critical temperature 
keeps its steady decrease with $B$, we require simulations
with an Euclidean time compactification length around 2 fm or larger,
meaning that, in order to reach lattice spacings
below 0.1 fm, we need to perform simulations on lattices with a 
large number of sites in the temporal direction (a few tens).
That sets by itself a strong limitation to the explorable lattices,
in particular regarding the aspect ratios (ratio of the 
spatial to the Euclidean time lattice extents) which are reasonably 
affordable, given the available computational resources. 
In particular regarding the 
approach to the thermodynamical limit,
our results should be considered as exploratory,
but nevertheless providing an already consistent
and clear picture, which claims for future 
investigations and refinements.

The paper is organized as follows. In Section~\ref{methods} we discuss
the lattice discretization of the theory and other technical details 
regarding the implementation of the magnetic background field and 
the physical observables explored in our investigation.
In Section~\ref{results} we present and discuss our numerical results.
Finally, in Section~\ref{conclusions} we draw our conclusions 
and discuss future perspectives.

\section{Numerical Methods}
\label{methods}

As in Refs.~\cite{parrot,gabbiano}, 
we consider a discretization of $N_f = 2+1$ QCD 
based on the tree-level improved Symanzik pure gauge action~\cite{weisz,curci} 
and on stout rooted staggered fermions~\cite{kogut-susskind,morning},
i.e.~on the following partition function
\begin{equation}
  Z=\int{[DU]}\,e^{-S_{YM}}\prod_{f=u,d,s} \det{(M_{st}^f)}^\frac{1}{4},
\end{equation}
where $[DU]$ is the Haar measure for gauge links, 
$f$ is the flavor index, and the fermion matrix and the gauge action
are respectively
\beq
  {M^f_{st\ }}_{ij} &=& \hat m_f\delta_{ij}+\sum_{\nu=1}^{4}\frac{\eta_{i;\nu}}{2}
\left(U^{(2)}_{i;\nu}\delta_{i\, j-\hat{\nu}} - U^{(2)\dagger}_{i-\hat{\nu};\nu}\delta_{i\,j+\hat{\nu}}\right) \nonumber \\
\label{dirac_operator}
  S_{YM} &=& -\frac{\beta}{3} \, \sum_{\substack{i \\ \mu\ne\nu}}\left(\frac{5}{6}W^{1\times1}_{i,\mu\nu}-\frac{1}{12}W^{1\times2}_{i,\mu\nu}\right)
\eeq
with periodic (antiperiodic) boundary conditions in the Euclidean temporal direction
for bosonic (fermionic) fields, in order to reproduce thermal conditions.
There, $i,j$ and $\mu,\nu$ are respectively lattice sites and directions, 
while $\beta$ is the inverse gauge coupling, $a$ is the lattice spacing
and $\hat m_f = a m_f$ are the dimensionless bare quark masses. 
The $\eta_{i;\nu}$ are the staggered quark phases, 
$U^{(2)}_{i;\nu}$ is the two-times stout smeared link 
(with isotropic smearing parameter $\rho=0.15$), while
$W^{1\times \cdot}_{i,\mu\nu}$s are the real parts of the trace 
of the link products along the $1\times1$ and $1\times2$ rectangular closed path, 
respectively. 

Bare quark masses and the gauge coupling values have been tuned in order to move 
on a line of constant physics, which reproduces 
experimental results for hadronic observables,
based on the determinations reported in Refs.~\cite{tcwup1,befjkkrs,physline3}.
In particular, as in Ref.~\cite{parrot}, 
we have considered three different lattice spacings, 
$a \simeq 0.057, 0.086$ and 0.114~fm; for each lattice spacing the 
physical temperature of the system, which is equal to 
the inverse of the Euclidean temporal extension, $T = 1 / (N_t a)$,
has been tuned by changing the number of temporal lattice sites $N_t$ at fixed $a$.
Such a fixed scale approach to thermodynamics has the drawback of not allowing
for a fine tuning of the physical temperature, however it has many advantages at 
the same time, since it simplifies both the renormalization of physical observable and the 
continuum extrapolation at fixed physical values 
of the external background field, as we discuss in the following.

\subsection{External magnetic field}

In the lattice approach, the presence of an external 
magnetic background field can be translated in the introduction 
of additional $U(1)$ phases to the elementary parallel transporters
\begin{equation}
  U^{(2)}_{i;\mu} \to u_{i;\mu}^f U^{(2)}_{i;\mu} \, 
\end{equation}
which are kept constant, i.e.~no functional integration is performed
over them, and are different for the different flavors, depending 
on their electric charge.
In particular, considering
a uniform magnetic field $\vec{B}$ in the $\hat{z}$ direction
and the following gauge choice
\begin{equation}\label{four_potential}
  A_t=A_x= A_z= 0, \qquad A_y(x) = Bx \, ,
\end{equation}
a possible discretization on a periodic toroidal lattice
is the following
\begin{equation}
  u_{i;y}^f=e^{i a^2 q_f B \,i_x}, \qquad {u_{i;x}^f\vert}_{i_x=L_x}=e^{-ia^2q_fL_xBi_y},
\end{equation}
with all other $U(1)$ link  variables set to one,
where $L_i$ is the number of lattice sites
along direction $i$ and last condition
guarantees smoothness 
of the magnetic field across the 
$x$-boundary~\cite{tHooft:1979rtg,wiese,review}.
This choice leads to a constant magnetic field but for a single 
plaquette, which is pierced by an additional Dirac string and guarantees
a zero magnetic flux across the lattice torus; invisibility 
of that string leads to a quantization condition for $B$,
which is more compelling for the smallest quark charge $q_f = e/3$:
\begin{equation}
  q_f B=\frac{ 2 \pi b_z}{a^2L_xL_y} \implies
  e B=\frac{ 6 \pi b_z}{a^2L_xL_y}, \qquad b_z \in \mathbb{Z} \, .
\label{def_B}
\end{equation}
The external field leads to 
additional discretization errors. Since the magnetic field acts on 
the system through the gauge
invariant $U(1)$ phase factors
that dynamical quarks pick up going through 
closed loops on the lattice, the phase factor for the smallest non-trivial 
loop (a plaquette in the $xy$ plane)
\beq
\exp\left(i q_f B a^2\right) = \exp\left(i \frac{6 \pi b_z}{L_x L_y} \frac{q_f}{e} \right) \, 
\eeq
must be much smaller than $2 \pi$, hence
\beq
\frac{2 b_z}{L_x L_y} \ll 1 \, 
\label{uvemcondition}
\eeq
where we have considered the up quark, for which discretization errors are larger;
all that sets a UV cut-off for the largest field explorable 
for a given lattice spacing, 
$e B \leq 2 \pi / a^2$. For the coarsest lattice studied in this 
study, $a \simeq 0.114$~fm, the cut-off is around 20~GeV$^2$, which is not 
too far from $eB = 9$~GeV$^2$: this is at the origin of 
sizable discretization effects observed for this value 
of the magnetic field, which disappear only after
a proper continuum extrapolation~\cite{parrot}.

\subsection{Observables}

The determination of the (pseudo)critical temperature 
$T_c$ will be based on the analysis of the renormalized chiral condensate
and of the susceptibility of the strange quark number, which are two 
standard observables used for the same purposes in previous studies.

The $f$-flavor condensate is defined as
\begin{equation}
\label{def_chircond1}
\langle\bar{\psi}\psi\rangle_f \hspace{-1pt} = \hspace{-1pt} \frac{\partial}{\partial m_f} 
\hspace{-2pt} \left( \hspace{-2pt} \frac{T}{V_s} \log Z \hspace{-2pt} \right) \hspace{-2pt}
= \frac{1}{4 a^3 L_s^3 N_t} \left\langle \tr (M^f_{st})^{-1}\right\rangle
\end{equation} 
where $V_s$ is the spatial volume and the trace of the inverse fermion matrix is determined 
configuration by configuration, as usual, by means of noisy estimators.
The condensate is affected by both additive and
multiplicative renormalizations, which can be subtracted
following the prescription of Ref.~\cite{Endrodi2011}
\begin{equation}
\label{eq:ren_pres_wupp}
\langle \bar{\psi}\psi\rangle_f^r (B,T)\hspace{-2pt}=\hspace{-2pt}
\frac{m_{f}}{m_{\pi}^2 F_\pi^2} \hspace{-2pt} \left(\langle\bar{\psi}\psi\rangle_{f}(B,T)
\hspace{-1pt} - \hspace{-1pt} \langle\bar{\psi}\psi\rangle_{f}(0,0)\right) 
\end{equation}
The zero-$T$ subtraction, which is performed at fixed UV cut-off, 
eliminates additive divergences, while
multiplication by the bare quark mass $m_f$ takes care of multiplicative
ones.

In the following we will show results for the sum of up and down 
contributions, i.e.~the renormalized  
light quark condensate
$\Sigma^r_l(B,T)$.
The behavior of $\Sigma^r_l(B,T)$ will be monitored to locate $T_c$, 
looking for its inflection point in the region where it drops towards
zero.
Just for the purpose of a finite size scaling analysis
around the transition, we will consider also the 
unrenormalized disconnected chiral susceptibility
\begin{equation}
\label{susc_sconn}
\chi_{\bar\psi\psi,f}^{disc} \equiv \frac{1}{16 L_s^3 N_t}
\left[\langle (\tr M_f^{-1})^2\rangle-\langle \tr M_f^{-1}\rangle^2 \right] \, .
\end{equation}

The dimensionless susceptibility of the strange quark number is instead defined
as follows (with $f = s$):
\beq
\chi_f &\equiv& \frac{1}{T^2}
\hspace{-1pt} \frac{\partial}{\partial \mu_f^2} 
\hspace{-2pt} \left( \hspace{-2pt} \frac{T}{V_s} \log Z \hspace{-2pt} \right) \hspace{-2pt}\\
&=& \frac{N_t}{4 L_s^3} 
\left\langle \tr \left[ M_f^{-1} \partial^2_{a \mu_f} M_f - 
\left(M_f^{-1} \partial_{a \mu_f} M_f \right)^2 
\right]\right\rangle \nonumber
\eeq
where $\mu_f$ is the quark chemical potential and, in the last line, only
terms which do not vanish at $\mu_f = 0$ have been left in\footnote{Considering
the standard introduction of the chemical potential on the lattice, where temporal 
gauge links in the forward/backward temporal direction 
get multiplied by a factor $\exp(\pm a \mu_f)$, the derivatives
$\partial_{a \mu_f} M_f$ and 
$\partial^2_{a \mu_f} M_f $ correspond to just the temporal part 
of the Dirac operator, with an additional minus 
for each derivative in the backward propagation.}.

\section{Numerical Results}
\label{results}

Most of our simulations have been carried out at three
different lattice spacings, 
$a \simeq 0.057, 0.086$ and 0.114~fm, keeping the spatial size
fixed at $a L_s \simeq 2.75$~fm and varying the 
temporal lattice size $N_t$ in order to change the temperature.
For reasons to be discussed below, 
results at the finest lattice spacing are only available for 
$eB = 4$~GeV$^2$.
In this setup the magnetic field, according to Eq.~(\ref{def_B}),
is kept fixed in physical units
by just using the same number of quanta $b_z$ for every 
lattice spacing: that makes the 
continuum extrapolation much easier. 

In particular, we 
have fixed $b_z = 41$ and $b_z = 93$ respectively for 4 and 9~GeV$^2$.
Larger spatial sizes, up to $\sim 4$~fm, have been explored in a few cases, in order 
to check the impact of finite size effects, or to perform a finite size scaling analysis
around the transition: in those cases, $b_z$ has been increased
accordingly in order to keep $eB$ fixed, see Eq.~(\ref{def_B}).
Additional simulations, needed for zero temperature
subtractions or normalization, have been performed 
for $eB = 0, 4$ and 9~GeV$^2$
on lattices with a temporal extension of around $5.5$~fm,
which is large enough to be considered as a good approximation
for $T \simeq 0$\footnote{Given the relatively low 
temperatures explored in this study, this is not a trivial
statement. Actually, our reference ``zero temperature'' lattice
corresponds to $T \simeq 36$~MeV, which is well below the 
explored values of $T$ and deep in the confined region,
at least for the present values of $eB$.}.

Monte-Carlo sampling of gauge configurations has been
performed based on a Rational Hybrid Monte-Carlo (RHMC) algorithm running on 
GPUs~\cite{openacc1,openacc2}. For each simulation we performed $O(10^3)$ RHMC trajectories of unit length, 
taking measures every $5$ trajectories. 

\subsection{The finite temperature transition at large $eB$}

\begin{figure}[t!]
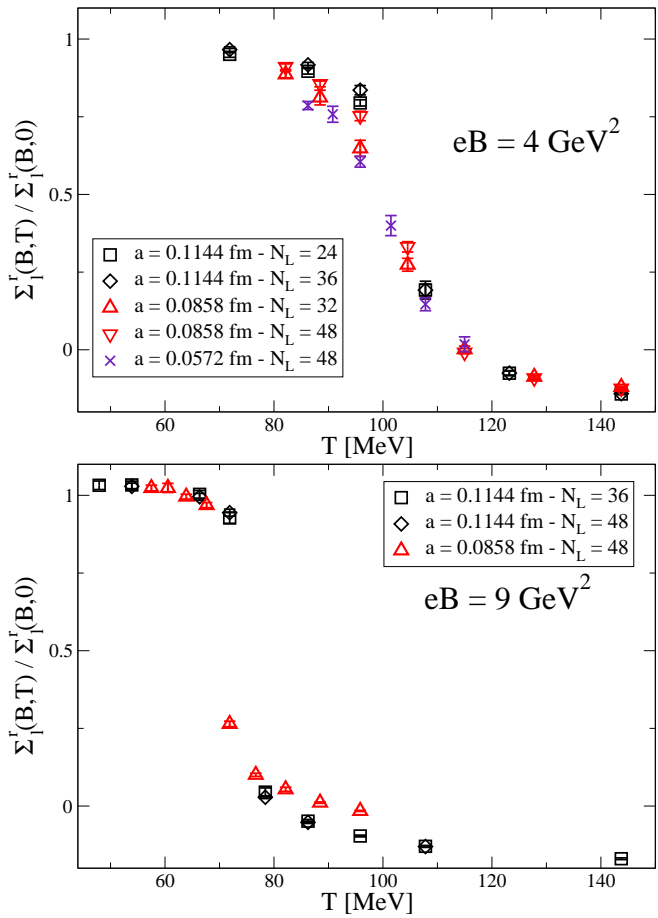

\includegraphics*[width=1\columnwidth]{plot_chiral_cond_4GeV_self_renorm.eps}\\
\includegraphics*[width=1\columnwidth]{plot_chiral_cond_9GeV_self_renorm.eps}\\
\caption{Renormalized chiral condensates, divided by their values at $T = 0$,
obtained for $eB = 4$~GeV$^2$ (top) and $eB = 9$~GeV$^2$ (bottom) 
for various lattice spacings and spatial extensions.
The drop of the pseudocritical temperature is clearly visible, as well
as the appearance of a well defined gap at the larger value of $eB$.}
\label{fig_condensates} 
\end{figure}

In Fig.~\ref{fig_condensates} we show the 
renormalized light condensate $\Sigma_l^r(B,T)$
as a function of $T$ for the two explored values of $eB$
and for various lattice spacings and spatial extensions.
Results have been normalized by those 
obtained for the same values of $B$ at $T \simeq 0$:
that suppresses much of the UV cut-off dependence 
already observed for $eB = 9$~GeV$^2$ at $T = 0$
in Ref.~\cite{parrot}. 

A residual UV cut-off,
as well as a finite size dependence, is visible around 
the transition, however that does not obscure the 
main message emerging from Fig.~\ref{fig_condensates}. 
$T_c$ is around 
100~MeV for $eB = 4$~GeV$^2$ and
drops below 80~MeV for $eB = 9$~GeV$^2$.
Moreover, one observes a significant strengthening of the transition,
which seems to become strong first order, with a large gap
in the chiral condensate,
at the larger value of $eB$.

Some considerations should be made about the possible weaknesses
of our results. We have been forced to work 
with aspect ratios  $L_s / N_t$ around 2, which is 
marginally compatible with a reliable study of thermodynamics, 
by some converging constraints: the fact that 
the range of physically relevant temperatures turns out to be 
lower than expected from previous lattice studies~\cite{Endrodi:2015oba},
and the need for 
lattice spacings fine enough to support the explored values of $eB$,
all that combined with a limited budget
of available computational resources. This is also the reason 
we do not have results available for the finest lattice spacing
at $eB = 9$~GeV$^2$, since in that case, without a significant 
increase of $L_s$, the aspect ratio 
would have been close to 1 around the transition.

Nevertheless, the main results depicted above
do not seem to be much affected by such weaknesses.
The dependence on the finite spatial size is visible around the transition
but is not significant. The value of $T_c$ at 
$eB = 9$~GeV$^2$, where only two lattice spacings are available,
seems to decrease even more when moving from the coarser
to the finer lattice, while the transition 
is sharp and seemingly strong first order in both cases.

\begin{figure}[t!]
\includegraphics*[width=1\columnwidth]{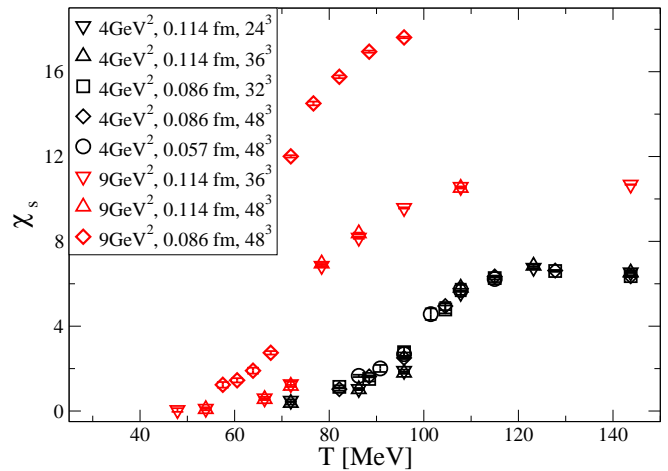}
\caption{Strange quark number susceptibility as a function 
of $T$ for both explored values of $eB$ and for
various lattice spacings and spatial extensions.
Significant UV cut-off effects are visible for
the larger magnetic field, however this does not affect
the conclusion for the appearance of a large gap at the transition 
in that case.}
\label{fig_strangesusc} 
\end{figure}

Similar conclusions are obtained by looking at results
for the strange quark susceptibility, which are reported in 
Fig.~\ref{fig_strangesusc}. The susceptibility raises 
in correspondence of the same temperatures at which 
the chiral condensate drops, and has a sudden jump,
suggesting a strong first order transition, for 
$eB = 9$~GeV$^2$. In the latter case, UV cut-off effects are clearly
visible and significant, even if they affect mostly the overall magnitude
of the susceptibility, and only marginally the location of $T_c$:
similar significant UV cut-off effects have been reported in Ref.~\cite{parrot},
at the same value of $eB$, for the chiral condensate, and can be ascribed 
to the rough discretization of such large magnetic field,
since up quarks pick an elementary phase around 
plaquettes which is large ($\sim 2 \pi /3$ and  $\sim \pi /3$,
respectively, for $a = 0.114$~fm and $a = 0.086$~fm).
We notice that the magnetic field induces a strong enhancement 
in quark number susceptibilities: similar observations have been 
reported in Ref.~\cite{Ding:2020pao}.
\\

\begin{figure}[t!]
\includegraphics*[width=1\columnwidth]{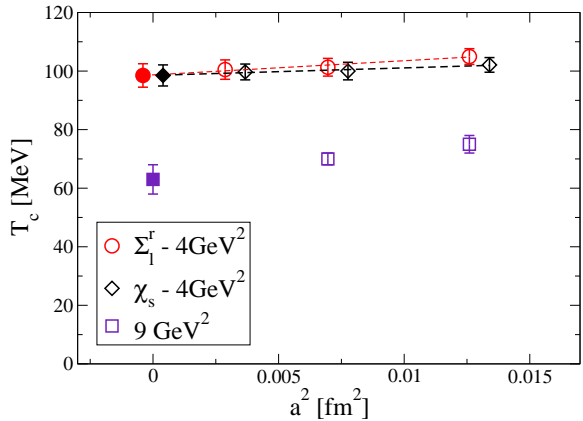}
\caption{Transition temperatures as a function of 
$a^2$, determined 
from the chiral condensate and from the strange
quark number susceptibility
at the two explored values of $eB$.
For $eB = 4$~GeV$^2$, $T_c$ has been determined
by fitting the inflection point of $\Sigma_l^r$ or $\chi_s$.
For $eB = 9$~GeV$^2$, instead, the determination
is obtained from the sharp jump observed for both quantities,
with an uncertainty given by the half-difference
of the temperatures on the two sides of the jump.
A tentative continuum extrapolation of $T_c$, assuming $O(a^2)$ corrections, 
is reported in both cases, however for $eB = 9$~GeV$^2$ this is not 
even a fit, since only two lattice spacings are available.}
\label{fig_tcextrap} 
\end{figure}

Results obtained for $T_c$ from both 
observables are shown as a function of $a^2$
in Fig.~\ref{fig_tcextrap}.
For $eB = 4$~GeV$^2$ $T_c$ has been determined
by fitting the inflection point of $\Sigma_l^r$ or $\chi_s$,
while for $eB = 9$~GeV$^2$ the determination
coincides with the midpoint of the two temperatures
where the sharp jump is observed, with an uncertainty given 
by their half-difference; a systematic uncertainty of around 2~\%, related
to the determination the lattice spacing~\cite{tcwup1,befjkkrs,physline3}, 
should be considered in both cases.
A tentative continuum extrapolation of $T_c$, assuming $O(a^2)$ corrections, 
is also reported for $eB = 4$~GeV$^2$, leading to 
$T_c (eB = 4 {\rm GeV}^2) = (98 \pm 3)$~MeV, 
while for $eB = 9$~GeV$^2$ we do not
have enough degrees of freedom even for a linear fit. In the latter case,
given the two available lattice spacings and all other systematic uncertainties,
we believe that a safe and conservative estimate for the continuum extrapolated
temperature is $T_c (eB = 9 {\rm GeV}^2) = (63 \pm 5)$~MeV.
\\

\begin{figure}[t!]
\includegraphics*[width=1\columnwidth]{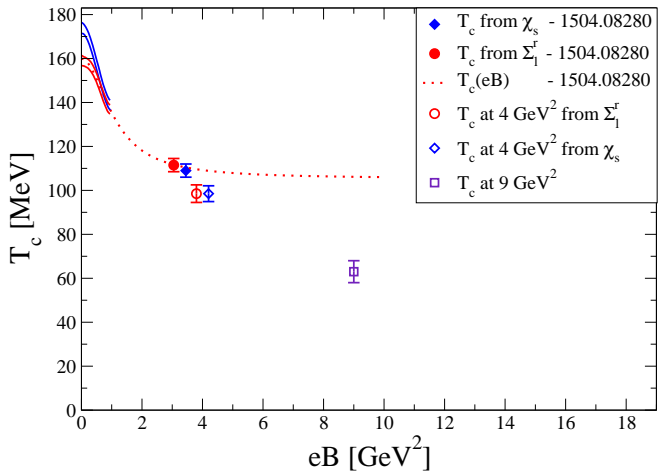}
\caption{We draw a first sketch of the 
updated version of the QCD phase diagram 
in a magnetic field, where our continuum extrapolated 
determinations of 
$T_c$ at $eB = 4$ and 9~GeV$^2$ are plotted together
with previous lattice 
determinations as well as tentative 
extrapolations (dotted line).
The blue and red bands in the small field region are continuum
extrapolations from Ref.~\cite{reg0}, obtained 
respectively from the 
strange quark number susceptibility and the quark condensate.
The determinations of Ref.~\cite{Endrodi:2015oba} 
are not extrapolated to the continuum limit,
which may account for their values seeming a bit higher.
}
\label{fig_diagram1} 
\end{figure}

We have put the two critical temperatures, together with previous results
available in the literature, in order to draw a first tentative sketch of the updated QCD phase diagram 
in a magnetic field, which is reported in Fig.~\ref{fig_diagram1}.
A first observation is that our results, which are consistent with 
all previous direct lattice determinations, point to a steady decrease 
of $T_c$ even in the large field region, contrary to a much smoother
approach to the infinite $B$ limit reported in the investigation of 
Ref.~\cite{Endrodi:2015oba},
which however was based on an effective description
of QCD at large $eB$ in terms of an anisotropic 
pure gauge theory~\cite{Miransky:2002rp,Miransky:2015ava}.
The second observation is that our results strongly suggest 
the presence of a strong first order transition, with a 
critical endpoint along the line which continously connect
$T_c(eB = 0)$ with $T_c(eB = 9~{\rm GeV}^2)$.
The presence of a first order transition at large $eB$ was predicted
in previous literature, with an 
estimate for the critical endpoint, based on an extrapolation, at
$eB_c = 10(2)$~GeV$^2$~\cite{Endrodi:2015oba};
our results suggest, for the first time from a direct lattice determination,
that the critical point is located somewhere 
in the middle between 
$eB = 4~{\rm GeV}^2$ and 
$eB = 9~{\rm GeV}^2$.
Such a conclusion however requires some deepening of our investigation, based 
on a finite size scaling analysis, in order to assess that
at $eB = 9~{\rm GeV}^2$ the transition is indeed first order:
this is done in the following subsection.

\subsection{Finite size scaling around the critical temperature}

The fixed UV cut-off approach we have followed till now 
allows only for a discrete set of temperatures; as 
a consequence, a large jump in some observables somewhere 
is only suggestive of a first order transition, 
but does not necessarily imply
it. Smoking guns would be instead 
the presence of metastable histories, double peak distributions
and a proper finite size scaling (FSS) analysis, which however require 
a fine tuning of the temperature around the transition point.

Therefore, in order to clarify this aspect, we have decided
to give up our fixed cut-off approach for a set of 
dedicated simulations. In particular, we have chosen 
one of the two simulation points at
$eB = 9~{\rm GeV}^2$ adiacent to the jump, 
taking it as the starting point for a 
temperature scan where $N_t$ is kept fixed 
and $T$ is changed by tuning the lattice spacing 
through the bare parameters. 

As a further variation, the lattice spacing
has been changed by tuning just the inverse 
gauge coupling $\beta$, which enters the 
pure gauge action, and not the bare quark masses,
which enter the fermion determinant. That 
simplifies the FSS analysis, allowing for an easy 
application of the multi-histogram method, in particular
without the need for a costly reweighting of the fermion determinant.
On the other hand, that has the drawback of moving us away 
from the physical line, however it should be clear that
this is not a relevant aspect: we are just doing a fine tuning, 
with the purpose of crossing the critical surface somewhere 
close to the starting point and test if it is first order or not;
since the presence of a first order transition, i.e.~of a gap
in physical observables, is stable under small variations
of the parameters, we will obtain a valid and clear-cut 
answer anyway.

\begin{figure}[t!]
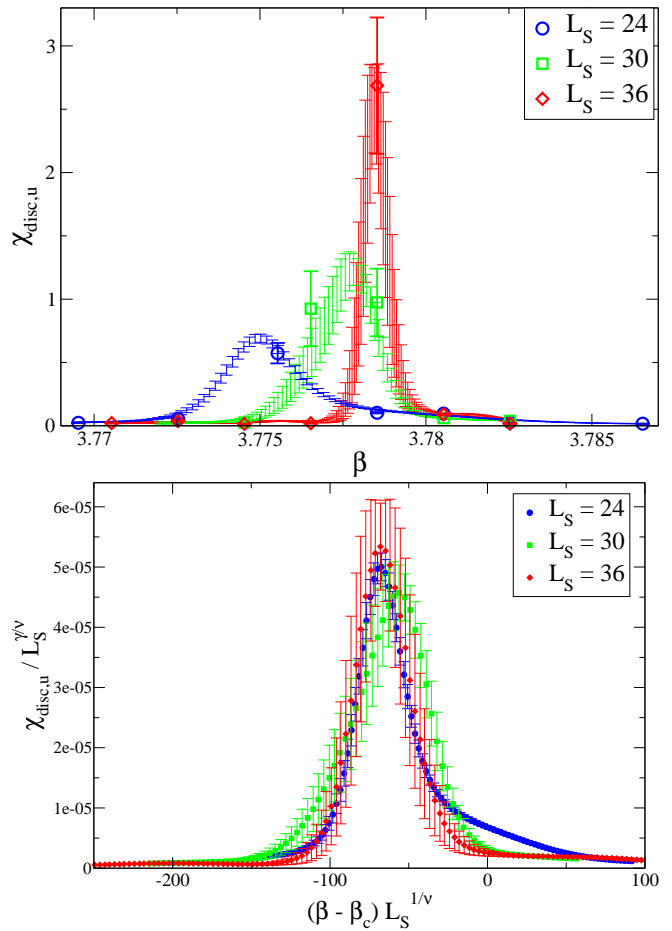

\includegraphics*[width=1\columnwidth]{plot_multihistogram_9GeV2}\\
\includegraphics*[width=1\columnwidth]{plot_multihistogram_rescaled_9GeV2}
\caption{FSS analysis of the unrenormalized disconnected chiral
susceptibility of the up quark. Data have been obtained on lattices
with $N_t = 22$, fixing $b_z = 93, 145, 209$ respectively 
for $L_s = 24,30,36$, so as to keep $eB$ constant as the thermodynamical
limit is approached; the inverse gauge coupling
$\beta $ has been tuned while keeping the bare 
quark masses fixed at 
$a m_s = 0.0457$ and $a m_{u/d} = 0.00162$.  
The FSS ansatz has been checked (lower figure)
by fixing $\nu = 1/3$ and $\gamma = 1$, as expected around a first
order transition, with $\beta_c \simeq 3.780$. 
}
\label{fig_multihist} 
\end{figure}

In order to make the computational effort of the FSS analysis 
affordable, we worked on the coarsest lattice; on the other
hand, the jumps observed in the chiral condensate and in the 
strange quark number susceptibility suggest that the transition
does not weaken going towards the continuum limit. 
As a starting point, we have chosen the $N_t = 22$ lattice
at $\beta = 3.787$, $a m_s = 0.0457$ and 
$a m_{u/d} = 0.00162$, 
which corresponds to $T \simeq 78.5$~MeV
and is the first point on the upper side of the transition,
and changed $\beta$ downwards, so as to increase the lattice 
spacing and decrease $T$, till we have crossed the transition.
This has been repeated for three different spatial
sizes, $L_s = 24, 30$ and 36.

\begin{figure}[t!]
\includegraphics*[width=1\columnwidth]{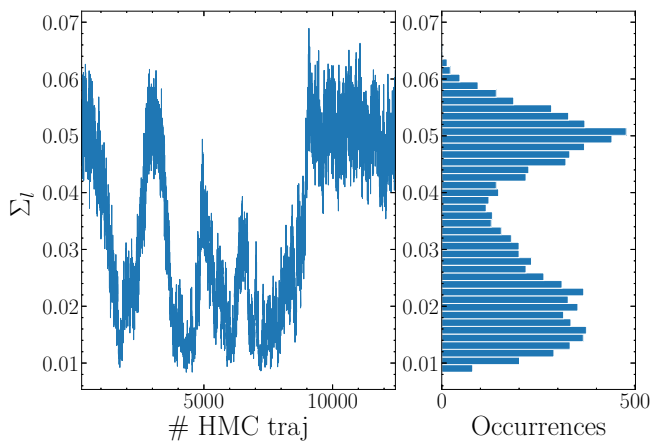}
\caption{MC history and distribution of the light quark condensate 
on the $24^3 \times 22$ lattice at $\beta = 3.7755$,
$a m_s = 0.0457$, $a m_{u/d} = 0.00162$ and $b_z = 93$.
The bistability and the corresponding double 
peak distribution are clearly visible.}
\label{fig_distr_chiral} 
\end{figure}

As a first result, in Fig.~\ref{fig_multihist} we show 
the disconnected and unrenormalized\footnote{Since we
want to explore the critical behavior of the 
chiral susceptibility as the thermodynamical 
limit is approached, looking at just the disconnected
part, which is expected to diverge itself at a 
genuine transition, is enough. For the same reason,
the subtraction of regular (at fixed UV cut-off) renormalization
constants is irrelevant to our purposes.}
chiral susceptibility $\chi_{disc,u}$ of the up quark 
(similar results are obtained for the down quark).
Results clearly show that the susceptibility increases
with the volume and that data collapse onto each other 
according to the following FSS ansatz ($\phi$ is an
unknown scaling function)
\beq
\frac{\chi_{disc,u} (L_s, \beta)}{L_s^{\gamma/\nu}} = 
\phi \left( (\beta - \beta_c) L_s^{1/\nu} \right) \, ,
\label{eq_fss}
\eeq
when $\nu$ and $\gamma$ are fixed to the expected effective first order
critical indexes for three spatial dimensions, i.e.~$\nu = 1/3$
and $\gamma = 1$. The critical value
of $\beta$ in Fig.~\ref{fig_multihist}, which optimizes
the collapse, is $\beta_c \simeq 3.780$.

As further evidence of the presence of a first transition, now we focus
on Monte-Carlo (MC) histories of some observables, looking for the 
presence of double peak distributions or metastable behaviors
around the transition. Fig.~\ref{fig_distr_chiral} shows
the MC history of the light chiral condensate, 
in units of HMC trajectories of unit length, 
on the $L_s = 24$ lattice at $\beta = 3.7755$: the history clearly oscillates
between two values, with a corresponding and well
defined double peak distribution. 

\begin{figure}[t!]
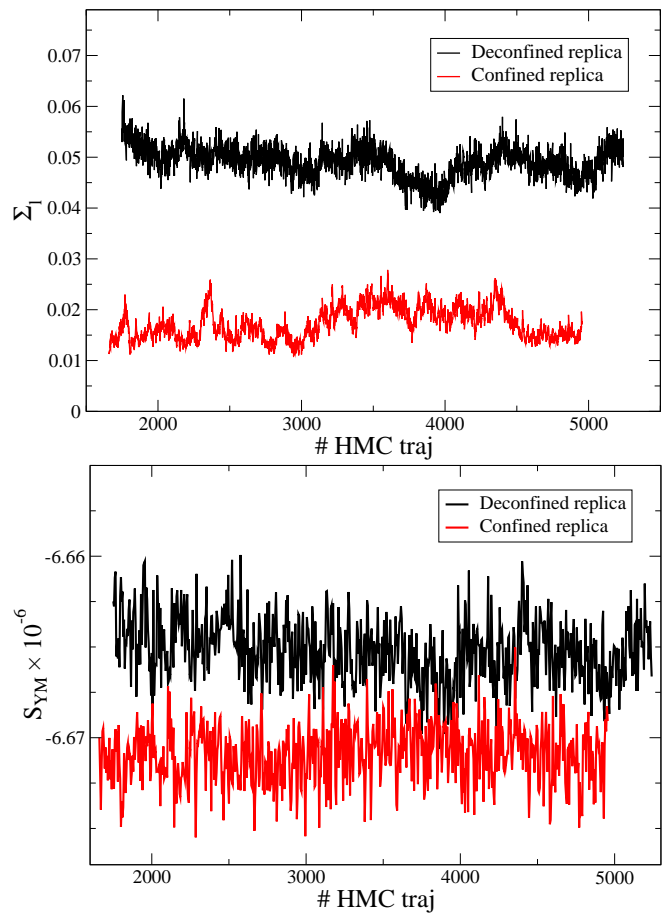

\includegraphics*[width=1\columnwidth]{plot_chiral_history_twin_pairs}\\
\includegraphics*[width=1\columnwidth]{plot_action_history_twin_pairs}
\caption{Two twin MC histories obtained on the 
$36^3 \times 22$ lattice at $\beta = 3.7785$,
$a m_s = 0.0457$, $a m_{u/d} = 0.00162$ and $b_z = 209$. 
The two runs have been started from different sides of the phase
transition, and keep staying in their starting phase for
the whole run, consisting of a few thousands of RHMC trajectories
of unit length. 
We show both the light quark condensate (top) and the pure 
gauge action (down).}
\label{fig_metastable} 
\end{figure}

As we move to a larger lattice, 
$L_s = 36$, the double peak distribution becomes so sharp
that the system is not able to easily tunnel from one phase
to the other in a reasonable MC time. This is clear from 
Fig.~\ref{fig_metastable}, where we show the MC 
histories of two twin runs, performed with exactly the same
parameters but starting from different sides 
of the phase transition: the two runs 
keep staying in their phase for a few thousands of RHMC 
trajectories; moreover, in this case the
bistability is clearly visible also in the pure gauge action.

\subsection{Confining properties of the two phases}

Having clarified that the large $B$ region of the 
$B - T$ phase diagram is characterized by a well defined 
phase separation, a number of interesting questions 
emerge, regarding the properties and differences between
the two phases. It is not the purpose of the present
investigation to give a comprehensive answer to such questions,
however we would like to touch at least one aspect, which
has been already considered in some previous 
studies and regards the confining 
properties of the theory~\cite{parrot, Mizher:2010zb, mostgentle, strongmag0, strongmag1, tusso, screening}. 

It is known that, at zero temperature, the static quark-antiquark potential
become anisotropic, with a suppression of the 
string tension in the direction parallel to the magnetic field,
and an enhancement in the transverse 
directions~\cite{parrot, mostgentle, strongmag0, strongmag1, tusso}.
The longitudinal string tension $\sigma_L$ is suppressed by more 
than one order of magnitude at $eB = 9~{\rm GeV}^2$, with respect
to its value at $eB = 0$, while the transverse string tension 
$\sigma_T$ seems to saturate its increase at a value which is 
around 50\% higher that the $B = 0$ value~\cite{parrot}.
The possible existence of a critical magnetic field $B_c$
at $T = 0$, where the longitudinal string tension
vanishes, and what could happen at such a critical field, 
is still unclear~\cite{parrot}. On the other hand, studies at 
finite temperature and up to moderate values of 
the magnetic field~\cite{strongmag1,screening} have shown
that anisotropies in the static potential become 
less significant approaching the phase transition.

As a minimal, additional contribution to the investigation
of the confining properties in the $B - T$ plane, we decided 
to investigate the static quark-antiquark potential 
at a fixed value of the temperature, $T \simeq 86$~MeV, for the 
two different explored magnetic fields, $eB = 4$ and 9~GeV$^2$.
According to Fig.~\ref{fig_diagram1}, the two simulations points should 
lay on the two different sides of the transition line.
In this case we have decided to perform the investigation
on the finest lattice, whose size is $48^3 \times 40$.

In order to determine the static quark-antiquark potential,
similarly to Refs.~\cite{parrot,strongmag0,strongmag1}, 
we studied the Wilson loop $\left< \tr W(a\vec{n},an_t) \right>$
and its dependence on the Euclidean time $a n_t$, 
exploiting the relation
\begin{equation}
  \left< \tr W(a\vec{n},an_t) \right> \propto e^{-aV(a\vec{n})n_t},
\end{equation}
which holds for large enough $an_t$. In particular,
from previous equation one can derive
\begin{equation}\label{operative_aV}
  aV(a\vec{n})=\lim_{n_t\to\infty}\log\left({\frac{\left< \tr W(a\vec{n},an_t) \right>}{\left< \tr W(a\vec{n},a(n_t+1)) \right>}}\right) \, ,
\end{equation}
so that 
the potential at fixed $\vec{n}$ can be obtained by fitting to a constant the $\log$ in the RHS of Eq.~(\ref{operative_aV}) as a function
of $n_t$, at least in a suitable stability range.

The application of such prescription in the present finite temperature
context might seem not appropriate. Indeed, because of the limited 
Euclidean temporal extension, the static quark-antiquark potential
is usually extracted from Polyakov loop correlators.
However, on one hand such correlators turns out to be extremely noisy
in our case, beyond the limit of feasibility, because of the relatively low temperatures
considered in our investigation, which imply a large number of lattice sites
in the temporal direction. On the other hand, because of the same reason,
the temporal extension turns out be large enough 
($N_t = 40$ in our particular case) and marginally compatible
with an extraction of the potential also from Wilson loops. 
It is clear that one should be careful about possible systematic effects 
related to this compromise, however the results we are going to show 
are clear-cut enough to make such systematics less worrying.

\begin{figure}[t!]
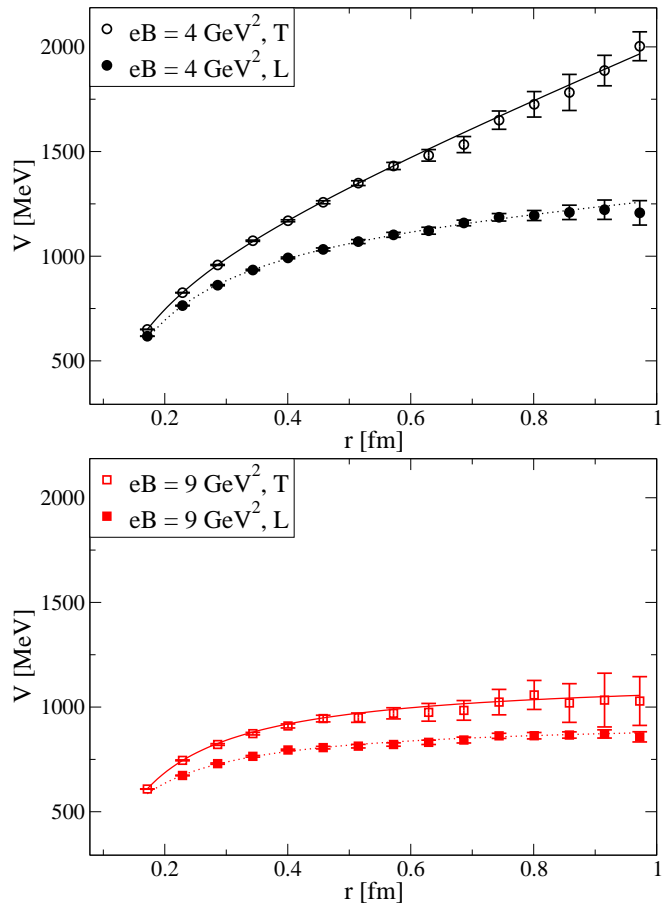

\includegraphics*[width=1\columnwidth]{plot_potentials_T40_4GeV}\\
\includegraphics*[width=1\columnwidth]{plot_potentials_T40_9GeV}
\caption{Static quark-antiquark potential, extracted from Wilson loops
computed on the $48^3 \times 40$ lattice at the two different values of 
the magnetic field and separately for the transverse (T)
and longitudinal (L) directions. The reported curves correspond to 
fit to the Cornell potential (for $eB = 4$~GeV$^2$) 
or to a purely Coulombic potential (for $eB = 9$~GeV$^2$).
} 
\label{fig_potential} 
\end{figure}

Results for the static quark-antiquark potential, computed
for the two different orientations and magnetic fields, are shown
in Fig.~\ref{fig_potential}: the different behavior in the two 
phases is particularly clear, also by eye, for the transverse direction,
where the linearly rising potential suddenly flattens moving 
from $4$ to $9$~GeV$^2$. In order to make a more quantitative
analysis, we have tried to fit data according to the Cornell
ansatz
\beq
V(r) = V_0 - \frac{\alpha}{r} + \sigma r
\label{eq_cornell}
\eeq
obtaining the following results. For $eB = 9$~GeV$^2$, data
are well fitted (with $\chi^2/{\rm d.o.f.} \lesssim 1$) 
by a purely Coulombic potential both in the trasverse and in the 
longitudinal direction; if one tries to include a non-zero 
$\sigma$, the fit returns negative values (for $\sigma_L$) or 
values compatible with zero within errors (for $\sigma_T$). 
For $eB = 4$~GeV$^2$, instead, a non-zero string tension is 
clearly needed in the transverse direction, with 
$\sqrt{\sigma_T} = 475(20)$~MeV, which is not far
from the $T = 0$ result obtained for the same
lattice spacing in Ref.~\cite{parrot}, 
$\sqrt{\sigma_T} \simeq 520$~MeV; for the longitudinal 
direction a full fit to Eq.~(\ref{eq_cornell})
returns $\sqrt{\sigma_L} = 215(20)$~MeV (which is 
close also in this case 
to the $T = 0$ result $\sqrt{\sigma_L} \simeq 240$~MeV~\cite{parrot}),
however one should consider that in this case reasonably good fits are 
obtained also assuming a purely Coulombic potential, if enough 
points are discarded at short distances.
\\

To summarize, present evidence is compatible, within numerical
uncertainties, with the transition from a strongly anisotropic
confined phase to a completely deconfined phase, in which the string
tension vanishes in all directions, as the critical line is
crossed. Such evidence should be supported by future studies, aimed at
assessing in a more precise way which string tension is vanishing or
not on both sides of the transition. In this respects, several
scenarios are plausible, including the possibile existence of an
intermediate phase in which $\sigma_L = 0$ but $\sigma_T \neq 0$, for
a subset of values of $B$ and $T$.

However, the sudden drop of the transverse string tension is a quite
clear and undoubtful phenomenon even now. In the simplest scenario,
one can assume that the critical temperature $T_c(B)$ continues its
drop as a function of $B$ until it hits the ground at some critical
magnetic field $B_c$. That would imply that, even at $T = 0$, there is
no transition to an anisotropically deconfined phase where $\sigma_L =
0$ and $\sigma_T \neq 0$, but rather a sudden transition to a
completely deconfined phase. Of course, even the assumption that
$T_c(B)$ hits the ground is not supported, at the present time, by any
other evidence.  \\

\section{Conclusions and Perspectives}
\label{conclusions}

The numerical results presented in this study 
update our understanding of the QCD phase diagram
in an external magnetic field in a substantial
way, bringing new facts and new speculations into the 
overall picture.
The main new results are that the (pseudo)critical temperature $T_c(B)$ 
continues its steady decrease as a function of $eB$, reaching 
values as a low as $60$~MeV for $eB$ of the order of $10$~GeV$^2$,
and that the crossover turns into a real first order
transition for large enough magnetic fields. 
The latter fact has been speculated for a long time:
in this paper we have provided first numerical
evidence based on lattice simulation of $N_f = 2+1$
QCD with physical quark masses. Moreover, we have
provided a first rough location of the critical 
endpoint $(B_E, T_E)$ of the first order line, 
with $4~{\rm GeV}^2 < e B_E < 9~{\rm GeV}^2$,
or alternatively $65~{\rm MeV} \lesssim T_E \lesssim 95~{\rm MeV}$). 

The existence and location of this critical endpoint
have many significant implications: from a phenomenological point of view, 
especially for the possible consequences stemming from a strong first order 
cosmological QCD transition, which could be observable 
even nowadays~\cite{Witten:1984rs,Applegate:1985qt}; from a theoretical
point of view, for a comparison with predictions from many effective
model studies~\cite{Cohen:2013zja,Moreira:2021ety,Ayala:2021nhx,Avancini:2012ee,Costa:2013zca,Bandyopadhyay:2020zte,Andersen:2021lnk,Mueller:2015fka}.

\begin{figure}[t!]
\begin{center}
\includegraphics*[width=1\columnwidth]{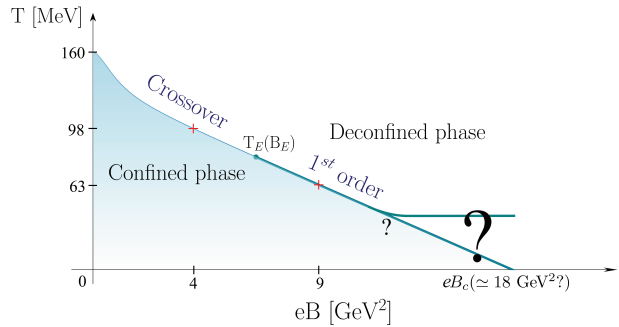}
\end{center}
\caption{Updated QCD Phase Diagram in an external magnetic field,
based on new facts and new speculations emerging from our numerical
investigations. The (pseudo)critical temperature $T_c(B)$ continues 
its steady drop as a function of $B$, and the transition switches
from a crossover to first order at a critical endpoint located 
in the range $4~{\rm GeV}^2 < e B_E < 9~{\rm GeV}^2$ (or alternatively
$65~{\rm MeV} < T_E < 95~{\rm MeV}$). The fact that $T_c(B)$ hits 
the ground at some finite critical magnetic field $B_c$ or not 
remains an open question for future studies.}
\label{fig_diagram2} 
\end{figure}

The new facts emerging from our investigation 
are reported in Fig.~\ref{fig_diagram2}, which represents
our present proposal for the QCD phase diagram. The proposal
contains also some question marks, concerning open issues and speculations,
that essentially regards the fate of $T_c(B)$ in 
the large $B$ limit. A naive linear extrapolation 
of present determinations of $T_c$ in the 
$B - T$ plane would imply
that $T_c$ vanishes for $e B_c \sim 20$~GeV$^2$:
does that really happen, 
and in that case would 20~GeV$^2$ be a natural
scale for $N_f = 2+1$ QCD?
Or does instead $T_c$ flatten for larger magnetic fields, approaching 
a finite value, or zero, only asymptotically? 

The issue will be likely solved by future studies, and is strictly 
correlated to the fate of the confining properties of 
the QCD vacuum in a strong magnetic field. 
Indeed, if any critical magnetic field exists at $T = 0$ where 
the confining properties of QCD get disrupted, this field likely 
coincides with the critical field where $T_c(B)$ hits the ground:
results from Ref.~\cite{parrot} indicate that such critical field,
if any, is larger than 9~GeV$^2$, and so do the finite $T$ results 
presented here. One interesting point emerging
from our study is that, as one crosses the critical line,
the string tension seems to vanish, within our present numerical
uncertainties, both in the longitudinal and in the transverse 
directions: if that applies down to $T = 0$, then one should not
expect any anisotropic deconfinement of the QCD vacuum at 
large fields, as hypothesized in Ref.~\cite{strongmag1},  
with the string tension vanishing only in the 
longitudinal direction, 
but rather a sudden quench of $\sigma$ 
in all directions at $B_c$.
\\

There is a number of relevant issues that should be refined
or investigated by future studies. 
First of all, one should consider that our study has been performed
with a compromise between the need for a fine lattice spacing,
in order to allow for a large magnetic field, and the need 
for large spatial sizes, in order to properly study thermodynamics.
The compromise, given the presently available computational
resources, has revealed to be not easy at all, essentially because
of the unexpectedly low temperatures reached by the critical line,
which forced us to work with low aspect ratios.
Even if we have shown that systematics related to the finite UV cut-off
and to the finite spatial size are reasonably under control, efforts
should be pursued in the future to improve on such systematics.

A more precise location of the critical 
endpoint $(B_E, T_E)$ could be achieved following 
different approaches. Since the first order transition 
at 9~GeV$^2$ seems to be quite strong, one could 
consider lower values of $eB$ and investigate how
the gap in physical observables changes along the transition line, 
trying to extrapolate
the point where it vanishes. Alternatively, one 
could start from the low $B$ region, trying to 
detect the critical behavior associated with the endpoint,
which is generally expected to be in the 3D-Ising universality class.

A future line of research should be dedicated to a precise
characterization of the 
properties and differences of 
the two phases along the first order transition. 
In this investigation
we have started a preliminary characterization of the confining 
properties, but many other relevant physical quantities 
should be considered, including a determination of the latent heat
along the first order line and
of the transport properties~\cite{conductivity,Finazzo:2016mhm}
in both phases.

Finally, present results, in particular those regarding 
the critical endpoint, should be put in the framework of  
a more general and multidimensional view of the QCD phase diagram, including 
a finite baryon chemical 
potential~\cite{Braguta:2019yci,Szymanski:2020stb,Buividovich:2021fsa,Abramchuk:2019lso,Ayala:2015lta,Ferrer:2016osp,Skokov:2011ib}, 
different number of light fermions~\cite{Kawaguchi:2021nsa} or 
a finite rotation~\cite{Yamamoto:2013zwa,Fukushima:2018grm,Chen:2021aiq,Yamamoto:2021oys}.

\acknowledgments

We thank M. Cardinali for collaboration in the early stages of this study.
Numerical simulations have been performed at the IT Center of the Pisa University and on the MARCONI and MARCONI100 machines at CINECA, based on the Project IscrB\_QGPSMF and on the 
agreement between INFN and CINECA (under projects INF20\_npqcd, INF21\_npqcd).
F.S. is supported by the Italian Ministry of University and Research (MUR) under grant PRIN20172LNEEZ and by INFN under GRANT73/CALAT.

\end{document}